\documentclass[amsmath,amssymb,twocolumn,prl,floatfix,showpacs,nofootinbib]{revtex4-2}
\usepackage{amsmath}
\usepackage{graphicx}
\usepackage{subfigure}
\usepackage{epstopdf}
\usepackage{color}
\usepackage{multirow}
\usepackage{setspace}
\usepackage{overpic}
\usepackage{amssymb}
\usepackage[colorlinks,urlcolor=blue, citecolor=blue,linkcolor=blue] {hyperref}
\usepackage{lineno}
\usepackage{bm}
\usepackage{rotating}
\usepackage[utf8]{inputenc}
\let\oldequation\equation
\let\oldendequation\endequation
\renewenvironment{equation}
 {\linenomathNonumbers\oldequation}
 {\oldendequation\endlinenomath}

\begin{document}

\title{\bf \boldmath
First Direct Measurement of the Absolute Branching Fraction of $\Sigma^+ \to \Lambda e^+ \nu_{e}$
}

\author{
\begin{small}
\begin{center}
M.~Ablikim$^{1}$, M.~N.~Achasov$^{13,b}$, P.~Adlarson$^{73}$, R.~Aliberti$^{34}$, A.~Amoroso$^{72A,72C}$, M.~R.~An$^{38}$, Q.~An$^{69,56}$, Y.~Bai$^{55}$, O.~Bakina$^{35}$, R.~Baldini Ferroli$^{28A}$
% ...
\\
\vspace{0.2cm}
(BESIII Collaboration)\\
\vspace{0.2cm} {\it
$^{1}$ Institute of High Energy Physics, Beijing 100049, People's Republic of China\\
% ...
$^{13}$ G.I. Budker Institute of Nuclear Physics SB RAS (BINP), Novosibirsk 630090, Russia\\
% ...
$^{28}$ INFN Laboratori Nazionali di Frascati , (A)INFN Laboratori Nazionali di Frascati, I-00044, Frascati, Italy; (B)INFN Sezione di Perugia, I-06100, Perugia, Italy; (C)University of Perugia, I-06100, Perugia, Italy\\
% ...
$^{34}$ Johannes Gutenberg University of Mainz, Johann-Joachim-Becher-Weg 45, D-55099 Mainz, Germany\\
$^{35}$ Joint Institute for Nuclear Research, 141980 Dubna, Moscow region, Russia\\
% ...
$^{38}$ Liaoning Normal University, Dalian 116029, People's Republic of China\\
% ...
$^{55}$ Southeast University, Nanjing 211100, People's Republic of China\\
$^{56}$ State Key Laboratory of Particle Detection and Electronics, Beijing 100049, Hefei 230026, People's Republic of China\\
% ...
$^{69}$ University of Science and Technology of China, Hefei 230026, People's Republic of China\\
% ...
$^{72}$ University of Turin and INFN, (A)University of Turin, I-10125, Turin, Italy; (B)University of Eastern Piedmont, I-15121, Alessandria, Italy; (C)INFN, I-10125, Turin, Italy\\
$^{73}$ Uppsala University, Box 516, SE-75120 Uppsala, Sweden\\
% ...
$^{b}$ Also at the Novosibirsk State University, Novosibirsk, 630090, Russia\\
% ...
}
\end{center}
\end{small}
}

\begin{abstract}
The first direct measurement of the absolute branching fraction of
$\Sigma^+ \to \Lambda e^+ \nu_{e}$ is reported based on an $e^+e^-$
annihilation sample of $(10087\pm44) \times 10^6$ $J/\psi$ events
collected with the BESIII detector at $\sqrt{s}=3.097$ GeV.  The
branching fraction is determined to be ${\mathcal B}(\Sigma^+ \to
\Lambda e^+ \nu_{e}) = [2.93\pm0.74(\rm stat) \pm 0.13(\rm
  syst)]\times 10^{-5}$, which is the most precise measurement
obtained in a single experiment to date and also the first result
obtained at a collider experiment.  Combining this result with the
world average of ${\mathcal B}(\Sigma^- \to \Lambda e^-
\bar{\nu}_{e})$ and the lifetimes of $\Sigma^{\pm}$, the ratio,
$\frac{\Gamma(\Sigma^- \to \Lambda e^- \bar{\nu}_{e})}{\Gamma(\Sigma^+
  \to \Lambda e^+ \nu_{e})}$, is determined to be $1.06 \pm 0.28$,
which is within 1.8 standard deviations of the value expected in the
absence of second-class currents that are forbidden in the Standard
Model.
\end{abstract}

\maketitle

\oddsidemargin  -0.2cm
\evensidemargin -0.2cm

In recent years, some experimental measurements, such as the lepton
flavor universality violation in $B$-meson decays~\cite{HFAG} and the
anomalous magnetic moment of the
muon~\cite{muon_g_2_BNL,muon_g_2_FNAL}, hint at physics beyond the
Standard Model (BSM).  More notably, the CDF collaboration has
recently reported a new precision measurement of the $W$-boson mass,
which deviates from the SM prediction by 7 standard
deviations~\cite{w_mass} and may again indicate the presence of BSM
physics.  In order to search for BSM physics, any independent test of
a SM prediction is valuable, such as the search for second-class
currents~\cite{theory_SW}, which are excluded from the SM.

Previous nuclear $\beta$ decay experiments gave contradictory
conclusions concerning the existence of second-class
currents. Positive signals have been reported by
F. P. Calaprice {\it et al.}~\cite{positive_scc_1}, in which the result requires a
second-class form factor comparable to the weak-magnetism term, and by
K. Sugimoto {\it et al.}~\cite{positive_scc_2, positive_scc_3, positive_scc_4}, where the
experimental results are in favor of the existence of the second-class
induced-tensor current.  On the other hand,
D. H. Wilkinson {\it et al.}~\cite{negative_scc_1, negative_scc_2, negative_scc_3,
  negative_scc_4} reported the absence of second-class currents.  In
addition, studies on hyperon semileptonic weak decays have been
performed, where the existence of second-class currents will yield a
nonzero pseudotensor term ($g_2$) in the matrix element of the
axial-vector current~\cite{hyperon_semileptonic}, and some of the
polarized $\Lambda$ experiments suggest a large
$g_2$~\cite{nonzero_g2}. Such a nonzero $g_2$ value could be caused by
flavor-SU(3)-symmetry-breaking
effects~\cite{nonzero_g2_su3_1,nonzero_g2_su3_2}, or it could reflect
the existence of a new interaction in the weak Hamiltonian involving
second-class currents.  To distinguish between genuine second-class
currents and flavor-SU(3)-symmetry-breaking effects, a unique
observable was first proposed by S. Weinberg~\cite{theory_SW} in 1958,
\begin{equation}
\label{eq_r}
R \equiv \frac{\Gamma(\Sigma^- \to \Lambda e^- \bar{\nu}_{e})}{\Gamma(\Sigma^+ \to \Lambda e^+ \nu_{e})}
=\frac{{\mathcal B}({\Sigma^- \to \Lambda e^- \bar{\nu}_{e}) \times \tau_{\Sigma^+}}}{{\mathcal B}(\Sigma^+ \to \Lambda e^+ \nu_{e}) \times \tau_{\Sigma^-}},
\end{equation}
where $\Gamma(\Sigma^- \to \Lambda e^- \bar{\nu}_{e})$ and
$\Gamma(\Sigma^+ \to \Lambda e^+ \nu_{e})$ are the partial widths of
the decay channels $\Sigma^- \to \Lambda e^- \bar{\nu}_{e}$ and
$\Sigma^+ \to \Lambda e^+ \nu_{e}$, respectively; ${\mathcal
  B}(\Sigma^- \to \Lambda e^- \bar{\nu}_{e})$ and ${\mathcal
  B}(\Sigma^+ \to \Lambda e^+ \nu_{e})$ are the branching fractions of
the above two channels; $\tau_{\Sigma^-}$ and $\tau_{\Sigma^+}$ are
the life times of $\Sigma^-$ and $\Sigma^+$, respectively.  In the
absence of second-class currents, this $R$ value should be just the
phase-space ratio for these two decays whether
flavor-SU(3)-symmetry-breaking effects exist or not, so any
experimental deviation from this expectation would be decisive
evidence for the existence of second-class currents~\cite{theory_SW}.
In 1960, the theoretical prediction of this $R$ value was calculated
to be 1.57 by T. D. Lee and C. N. Yang~\cite{theory_LY} on the basis
of no second-class currents.

In the late 1960s, the decay $\Sigma^{\pm}\to \Lambda e^{\pm} \nu$ was
studied at fixed-target
experiments~\cite{old_BF_result_1,old_BF_result_2,old_BF_result_3}.
All the experimentally determined $R$ values are consistent with the
theoretical calculation, i.e. 1.57, within uncertainty.  In the
following years, the precisions of $\tau_{\Sigma^{\pm}}$ and ${\mathcal
  B}(\Sigma^- \to \Lambda e^- \bar{\nu}_{e})$ measurements have been
improved significantly, but the current experimental uncertainty of
$R$~\cite{pdg2022} is still large due to the poor precision of
${\mathcal B}(\Sigma^+ \to \Lambda e^+ \nu_{e})$. Previously, there
were three experimental results of ${\mathcal B}(\Sigma^+ \to \Lambda
e^+ \nu_{e})$, and all of them are indirect
measurements~\cite{indirect_m} from fixed-target experiments, which
were performed more than 50 years
ago~\cite{old_BF_result_1,old_BF_result_2,old_BF_result_3}.  The
highest precision measurement of ${\mathcal B}(\Sigma^+ \to \Lambda
e^+ \nu_{e})$ was performed in 1969 based on 10 signal
events~\cite{old_BF_result_2}, which were selected from about 6 million bubble chamber
pictures. Therefore, a direct measurement of ${\mathcal B}(\Sigma^+ \to \Lambda e^+ \nu_{e})$ 
at a modern collider experiment with higher precision is crucial to provide
a more stringent test of the existence of second-class currents.

In this Letter, we report the first direct measurement of the absolute
branching fraction ${\mathcal B}(\Sigma^+ \to \Lambda e^+ \nu_{e})$,
by analyzing $\Sigma^+\bar{\Sigma}^-$ hyperon pairs in $(10087\pm44)
\times 10^6$~\cite{num_jpsi} $J/\psi$ events collected with the BESIII
detector at $\sqrt{s}=3.097$ GeV.  The $J/\psi$ meson decays into the
$\Sigma^+\bar{\Sigma}^-$ final state with a branching fraction of
$(1.07 \pm 0.04) \times 10^{-3}$~\cite{pdg2022}.  We use the
double-tag (DT) technique~\cite{DT}, which provides a clean and direct
branching fraction measurement without requiring knowledge of the
total number of $\Sigma^+\bar{\Sigma}^-$ events produced.  Based on
the measured absolute branching fraction of $\Sigma^+ \to
\Lambda e^+ \nu_{e}$, a higher precision $R$ is determined.
Throughout this Letter, the charge-conjugated ($c.c.$)
channel is always implied.

Details about the design and performance of the BESIII detector are
given in Ref.~\cite{BESIII}.  Simulated data samples produced with
{\sc Geant4}-based~\cite{geant4} Monte Carlo (MC) software, which
includes the geometric description of the BESIII detector~\cite{Huang:2022wuo} and the
detector response, are used to determine the detection efficiencies
and to estimate backgrounds. The simulation includes the beam energy
spread and initial state radiation in the $e^+e^-$ annihilations
modeled with the generator {\sc kkmc}~\cite{kkmc}.  For the simulation
of the signal channel $\Sigma^+ \to \Lambda e^+ \nu_{e}$, we use the
model reported in Ref.~\cite{wangruming}, which has been validated in
Ref.~\cite{pmunu}, and use the form factors obtained from the SU(3)
flavor parametrization and the Cabibbo theory, which are summarized in
Ref.~\cite{hyperon_semileptonic}.  The inclusive MC sample of generic
events includes both the production of the $J/\psi$ resonance and the
continuum processes incorporated in {\sc kkmc}~\cite{kkmc}.  All
particle decays are modeled with {\sc evtgen}~\cite{evtgen} using
branching fractions either taken from the Particle Data Group (PDG),
when available, or estimated with {\sc lundcharm}~\cite{lundcharm} for $J/\psi$ resonance.
Final state radiation from charged final state particles is
incorporated using {\sc photos}~\cite{photos}.


Using the DT technique, the branching fraction is obtained by
reconstructing signal $\Sigma^+ \to \Lambda e^+ \nu_{e}$ decay in
events with $\bar{\Sigma}^-$ decays reconstructed in their dominant
hadronic decay mode $\bar{\Sigma}^- \to \bar{p} \pi^0$, where the
$\Lambda$ and $\pi^0$ decays are reconstructed through $\Lambda \to p
\pi^-$ and $\pi^0 \to \gamma \gamma$, respectively.  If a
$\bar{\Sigma}^-$ hyperon is found, it is referred to as a single-tag
(ST) candidate.  An event in which a signal $\Sigma^+$ decay and a ST
$\bar{\Sigma}^-$ are simultaneously found is referred to as a DT
event. The branching fraction of the signal decay is given by

\begin{equation}
\label{eq_bf}
{\mathcal B_{\rm sig}}= \frac{N_{\rm DT}/\epsilon_{\rm DT}}{N_{\rm ST}/\epsilon_{\rm ST}} \times \frac{1}{{\mathcal B}(\Lambda \to p \pi^-)},
\end{equation}
where $N_{\rm DT}$ is the DT yield, $\epsilon_{\rm DT}$ is the DT
efficiency, ${\mathcal B}(\Lambda \to p \pi^-)$ is the
branching fraction of $\Lambda \to p \pi^-$, and $N_{\rm ST}$ and
$\epsilon_{\rm ST}$ are the ST yield and the ST efficiency,
respectively.

Charged tracks detected in the main drift chamber (MDC) are required
to be within a polar angle ($\theta$) range of
$|\rm{cos\theta}|<0.93$, where $\theta$ is defined with respect to the
$z$-axis, which is the symmetry axis of the MDC.  Photon candidates
are identified using showers in the electromagnetic calorimeter (EMC).
The deposited energy of each shower must be more than 25~MeV in the
barrel region ($|\cos \theta|< 0.80$) and more than 50~MeV in the end
cap region ($0.86 <|\cos \theta|< 0.92$).  To exclude showers that
originate from charged tracks, the angle subtended by the EMC shower
and the position of the closest charged track at the EMC must be
greater than 10 degrees as measured from the interaction point (IP).
To suppress electronic noise and showers unrelated to the event, the
difference between the EMC time and the event start time is required
to be within [0, 700]\,ns.

Particle identification~(PID) for charged tracks combines measurements
of the specific ionization energy loss in the MDC~(d$E$/d$x$) and the
flight time in the time-of-flight system to form likelihoods
$\mathcal{L}(h)~(h=p,K,\pi)$ for various hadron $h$ hypotheses.  A
track is identified as a proton when the proton hypothesis has the
greatest likelihood ($\mathcal{L}(p)>\mathcal{L}(K)$ and
$\mathcal{L}(p)>\mathcal{L}(\pi)$) and satisfies
$\mathcal{L}(p)>0.001$.  The $\pi^0$ candidates are reconstructed by
performing a one constraint (1C) kinematic fit on the chosen photon
pair, constraining the invariant mass of the photon pair to the
nominal $\pi^0$ mass, and the $\chi^2_{\rm 1C}$ of the kinematic fit
is required to be less than 25.  Events with at least one anti-proton
and one $\pi^0$ are assigned as ST
$\bar{\Sigma}^-$ candidates. 

Next, tagged $\bar{\Sigma}^-$ hyperons are selected using two
kinematic variables, the energy difference
\begin{equation}
\label{def_delE}
\Delta E_{\rm tag} \equiv E_{\rm beam} - E_{\bar{\Sigma}^-},
\end{equation}
and the beam-constrained mass
\begin{equation}
\label{def_mbc}
M_{\rm BC}^{\rm tag}c^2 \equiv \sqrt{E^{2}_{\rm beam}-|\vec{p}_{\bar{\Sigma}^-}c|^{2}},
\end{equation}
where $E_{\rm beam}$ is the beam energy, $\vec{p}_{\bar{\Sigma}^-}$
and $E_{\bar{\Sigma}^-}$ are the momentum and the energy of the
$\bar{\Sigma}^-$ candidate in the $e^+e^-$ rest frame, respectively.
If there are multiple combinations, the one giving the minimum
$|\Delta E_{\rm tag}|$ is retained for further analysis.  The tagged
$\bar{\Sigma}^-$ candidates are required to satisfy $\Delta E_{\rm
  tag}\in[-48,\, 52]$\,MeV.

\begin{figure}[htp]
  \centering
\includegraphics[width=1\linewidth]{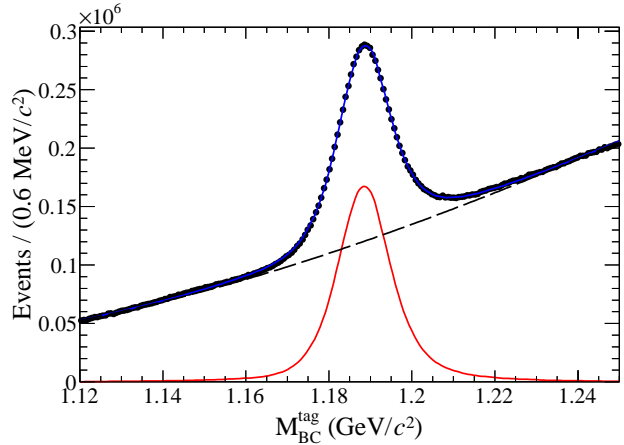}
  \caption{\small
Fit to the $M_{\rm BC}^{\rm tag}$ distribution of the ST
$\bar{\Sigma}^- + c.c.$ candidates.
Data are shown as dots with error bars.
The solid blue, solid red, and dashed black curves are the fit result, signal shape, and the background shape, respectively.}
\label{fig_STfit}
\end{figure}

The yield of ST $\bar{\Sigma}^-$ hyperons is obtained from a binned
extended maximum likelihood fit to the $M_{\rm BC}^{\rm tag}$
distribution of the surviving ST candidates, where the signal is
modeled by the MC-simulated signal shape convolved with a Gaussian
function to account for imperfect simulation of the
detector resolution, and the background is modeled by a third-order
Chebyshev function.  The
fit result is shown in Fig.~\ref{fig_STfit}, and the total ST
$\bar{\Sigma}^-$ yield is $N_{\rm ST}=4,693,360\pm4,339(\rm stat)$.

Candidate events for the $\Sigma^+ \to \Lambda e^+ \nu_{e}$, $\Lambda
\to p \pi^-$ decays are selected from the remaining tracks recoiling
against the ST $\bar{\Sigma}^-$ events in the mass
region $|M_{\rm BC}^{\rm tag}-m_{\bar{\Sigma}^-}|<0.049~ {\rm
  GeV}/c^{2}$, where $m_{\bar{\Sigma}^-}$ is the $\bar{\Sigma}^-$
nominal mass. The total number of charged tracks in the event is
required to be four ($N_{\rm Track}= 4$), and the selection criteria
for the additional charged tracks are the same as those used in the ST
selection. We further identify a charged track as an $e^+$ by
requiring the PID likelihoods, including also the EMC information, satisfy
$\mathcal{L}'(e)/(\mathcal{L}'(e)+\mathcal{L}'(\pi)+\mathcal{L}'(K))>0.8$,
where $\mathcal{L}'(e)$, $\mathcal{L}'(\pi)$, $\mathcal{L}'(K))$ are
likelihoods calculated based on the positron, pion and kaon
hypotheses, respectively.  A $\pi^-$ candidate is required to satisfy
$\mathcal{L}'(\pi) > \mathcal{L}'(K)$ and $\mathcal{L}'(\pi) >
\mathcal{L}'(e)$.  The other track is assumed to be a proton.  Each
$\Lambda$ candidate is reconstructed from a proton and a pion, which
are constrained to originate from a common vertex and are required to
fall in an invariant mass region $|M_{p \pi^-} - m_{\Lambda}|<$
10~MeV$/c^{2}$, where $m_{\Lambda}$ is the $\Lambda$ nominal
mass~\cite{pdg2022}.  The decay length of the $\Lambda$ candidate from the
IP is
required to be greater than twice the vertex resolution.

As the neutrino is not detected, we employ the kinematic variable
\begin{equation}
 M_{\rm miss}^2c^4\equiv E_{\rm miss}^2-{p}_{\rm miss}^2c^2,
\end{equation}
where $E_{\rm miss}$ and $p_{\rm miss}$ are the missing energy and
momentum carried by the neutrino, respectively.  $E_{\rm miss}$ is
calculated by
 \begin{equation}
 E_{\rm miss}=E_{\rm beam}-E_{\Lambda}-E_{e^+},
 \end{equation}
 where $E_{\Lambda}$ and $E_{e^+}$ are the energies of the
 $\Lambda$ and $e^+$ calculated in the $e^+e^-$ rest frame, respectively.  To obtain better resolution, we
 use the constrained $\Sigma^+$ momentum ($\vec{p}_{\Sigma^+}$) given
 by
  \begin{equation}
   \vec{p}_{\Sigma^+}=-\frac{\vec{p}_{\bar{\Sigma}^-}}{c|\vec{p}_{\bar{\Sigma}^-}|}\sqrt{E_{\rm beam}^2-m_{\Sigma^+}^2c^4},
 \label{eq_psigma}
 \end{equation}
 to calculate $p_{\rm miss}$
 \begin{equation}
   p_{\rm miss}=|\vec{p}_{\Sigma^+}-{\vec p}_{\Lambda}-{\vec p}_{e^+}|,
 \end{equation}
 where ${\vec p}_{\Lambda}$ and ${\vec p}_{e^+}$ are the momenta of $\Lambda$ and $e^+$
 calculated in the $e^+e^-$ rest frame, respectively.  For signal events,
 $M_{\rm miss}^2$ is expected to peak at zero.

For the signal candidates of $\Sigma^+ \to \Lambda e^+ \nu_{e}$ decay,
there are still some non-$\Sigma^+\bar{\Sigma}^-$ pair background
events.  Therefore, we require that the opening angle between the
anti-proton and $\pi^0$ in $\bar{\Sigma}^-$ rest frame satisfies ${\rm
Angle}(\bar{p},\pi^0) > 170^{\circ}$, and the momentum of the
anti-proton in the $\bar{\Sigma}^-$ rest frame satisfies
$p_{\bar{p}}^{\rm ST} \in [0.16,\,0.21]~{\rm GeV}/c$, where the
$\bar{\Sigma}^-$ rest frame is determined by making use of the fact
that $J/\psi \to \Sigma^+ \bar{\Sigma}^-$ is a two-body reaction.  To
further suppress backgrounds, we impose two extra requirements as
follows.  First, the
$\bar{\Sigma}^-\Lambda$ recoil mass calculated using
$M_{\bar{\Sigma}^-\Lambda}^{\rm recoil}+M_{p \pi^-}-m_{\Lambda}$ is required to be
greater than $-60$~MeV$/c^{2}$, which can suppress the contributions
from $\Sigma^+$ hadronic decay and other combinatorial background.
$M_{\bar{\Sigma}^-\Lambda}^{\rm recoil}$ is formed by
 \begin{equation}
  M_{\bar{\Sigma}^-\Lambda}^{\rm recoil}c^2=\sqrt{(E_{\bar{\Sigma}^-\Lambda}^{\rm recoil})^2-(p_{\bar{\Sigma}^-\Lambda}^{\rm recoil})^2c^2},
 \end{equation}
 where $E_{\bar{\Sigma}^-\Lambda}^{\rm recoil}$ and $p_{\bar{\Sigma}^-\Lambda}^{\rm recoil}$ 
 are the total energy and the total momentum of all recoiling particles against $\bar{\Sigma}^-\Lambda$ in an event, 
 which are calculated by
 \begin{equation}
E_{\bar{\Sigma}^-\Lambda}^{\rm recoil}=E_{\rm beam}-E_{\Lambda},
 \end{equation}
and 
 \begin{equation}
p_{\bar{\Sigma}^-\Lambda}^{\rm recoil}=|\vec{p}_{\Sigma^+}-{\vec p}_{\Lambda}|
 \end{equation}
respectively ($E_{\Lambda}$, $\vec{p}_{\Sigma^+}$, and ${\vec p}_{\Lambda}$ are calculated in the $e^+e^-$ rest frame). 
The variable $M_{\bar{\Sigma}^-\Lambda}^{\rm recoil}+M_{p \pi^-}-m_{\Lambda}$ can provide improved
resolution compared to $M_{\bar{\Sigma}^-\Lambda}^{\rm recoil}$~\cite{zc,zcs}.
Second, the decay of $\Sigma(1385)^+ \to \Lambda\pi^+$ could
contribute as background when the pion is wrongly identified as a
positron.  If we assign the $\pi^+$ mass to the $e^+$ candidates when
calculating the invariant mass of $\Lambda$$e^+$,
i.e. $M_{\Lambda(\pi^+ \to e^+)}$, the $\Sigma(1385)^+$ mass is
expected for the background.  Therefore, we can eliminate background
by only retaining the events with $M_{\Lambda(\pi^+ \to
  e^+)}<1.32~{\rm GeV}/c^{2}$.  The above four requirements can
effectively suppress the backgrounds, with a signal
efficiency loss of less than 1\% for each requirement.

\begin{figure}[htp]
  \centering
\includegraphics[width=1\linewidth]{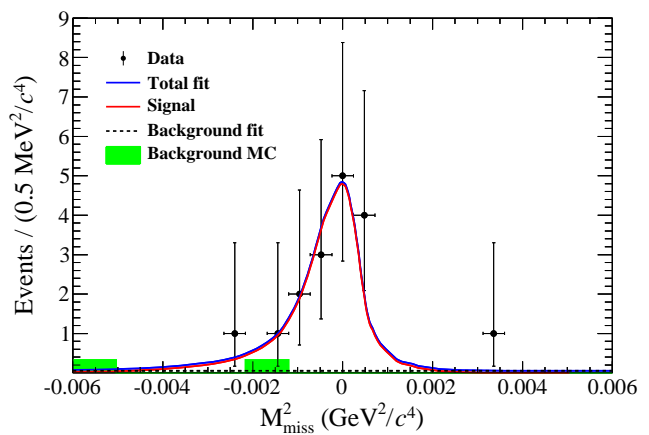}
  \caption{\small Fit to the $M_{\rm miss}^2$ distribution of the DT candidates. 
  Data are shown as dots with error bars.  
  The solid blue and red curves are the fit result and signal shape, respectively. 
  The black dashed curve is the background shape.
  The green-filled histogram is the background estimated from the inclusive MC sample.
    }
\label{fig_DTfit}
\end{figure}

The inclusive MC sample is analyzed using the TopoAna~\cite{topology} package to
study potential backgrounds.  After imposing the above selection
criteria, the contribution from the $\Sigma^+\bar{\Sigma}^-$ pair
background is negligible, and there are only a few
non-$\Sigma^+\bar{\Sigma}^-$ pair background events surviving, which
are shown as the green filled histogram in Fig.~\ref{fig_DTfit}.

To determine the signal yield, an unbinned extended maximum likelihood
fit is performed to the $M_{\rm miss}^2$ distribution.  The signal is
modeled by the MC-simulated signal shape convolved with a Gaussian
resolution function. The background is modeled by a flat background
shape.  The parameters of the Gaussian and all yields are free in the
fit. The fit to the data is shown in Fig.~\ref{fig_DTfit}, and $N_{\rm
  ST}$, $\epsilon_{\rm ST}$, $N_{\rm DT}$ and $\epsilon_{\rm DT}$ are
summarized in Table~\ref{tab_all}.

\begin{table}
\begin{center}
\caption{
\label{tab_all}
Values of $N_{\rm ST}$, $N_{\rm DT}$, $\epsilon_{\rm ST}$ and $\epsilon_{\rm DT}$ used for the branching fraction determination.
The uncertainties are statistical only.}
\begin{tabular}{cccc}
\hline\hline 
$N_{\rm ST}$&					$N_{\rm DT}$& 		$\epsilon_{\rm ST}$\,(\%)&	$\epsilon_{\rm DT}$\,(\%)\\
\hline
$4,693,360\pm4,339$&			$15.7\pm4.0$&		$41.48 \pm 0.03$&			$7.40\pm0.01$\\
\hline\hline
\end{tabular}
\end{center}
\end{table}

The systematic uncertainty due to the requirement for $N_{\rm Track}=
4$ (2.7\%) is studied with the control sample of ($J/\psi \to \Lambda
\bar{\Lambda}, \Lambda \to p \pi^{-}, \bar{\Lambda}\to \bar{p}
\pi^{+}$)~\cite{pmunu}, while the uncertainty for $\Lambda$
reconstruction (1.1\%) is studied with the control samples of ($J/\psi
\to \bar{p} K^+ \Lambda, \Lambda \to p \pi^{-}
+c.c.$) and $J/\psi \to p \bar{p} \pi^+
\pi^-$~\cite{lambda_recon}. The uncertainties due to the tracking
(1.0\%) and PID (2.3\%) for the positron are studied with the control
sample $e^+ e^- \to \gamma e^+ e^-$~\cite{sys_positron}.  For the
simulation of the signal MC model (0.6\%), it is estimated by varying
the input values of form factors~\cite{hyperon_semileptonic} by one
standard deviation.  The uncertainties due to the fits to the $M_{\rm
  miss}^2$ (0.3\%) and $M_{\rm BC}^{\rm tag}$ (1.8\%) distributions
are estimated by using alternative fit procedures, i.e. changing the
signal and background shapes for both of these fits and changing the
bin size for the fit to the $M_{\rm BC}^{\rm tag}$ distribution.  For
the branching fraction ${\mathcal B}(\Lambda \to p \pi^-)$, the
uncertainty is 0.8\%~\cite{pdg2022}.  The total systematic uncertainty
is estimated to be 4.4\% by adding all these uncertainties in
quadrature.

Finally, by using $N_{\rm ST}$, $\epsilon_{\rm ST}$, $N_{\rm DT}$ and
$\epsilon_{\rm DT}$ summarized in Table~\ref{tab_all} and the
well-measured ${\mathcal B}(\Lambda \to p \pi^-)= (63.9 \pm 0.5) \%$
from the PDG into Eq.~\ref{eq_bf}, the corresponding branching
fraction is determined to be ${\mathcal B}(\Sigma^+ \to \Lambda e^+
\nu_{e}) = (2.93\pm0.74\pm0.13)\times10^{-5}$, where the first
uncertainty is statistical and the second is systematic.  Combining
with the well-measured branching fraction of $\Sigma^- \to \Lambda e^-
\bar{\nu}_{e}$ [${\mathcal B}(\Sigma^- \to \Lambda e^- \bar{\nu}_{e})
  = (5.73\pm0.27) \times10^{-5}$] and the lifetimes of $\Sigma^{\pm}$
    [$\tau_{\Sigma^-}=(1.479\pm0.011)\times10^{-10}$ s and
      $\tau_{\Sigma^+}=(8.018\pm0.026)\times10^{-11}$
      s]~\cite{pdg2022}, we determine the $R$ value defined in
    Eq.~\ref{eq_r} to be $R = 1.06 \pm 0.28$.  This result is within
    1.8 standard deviations of the theoretical
    calculation~\cite{theory_LY} in the absence of second-class
    currents.

\begin{figure}[htp]
  \centering
\includegraphics[width=1\linewidth]{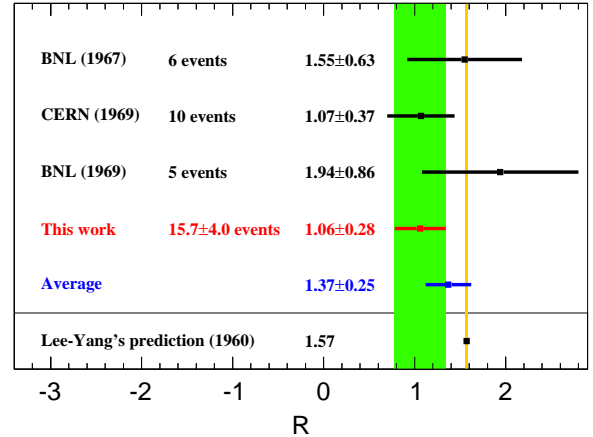}
  \caption{\small $R$ value
obtained in this work, compared to the recalculated values by previous fixed-target experiments 
(BNL (1967)~\cite{old_BF_result_3}, CERN (1969)~\cite{old_BF_result_2}, BNL (1969)~\cite{old_BF_result_1}), 
the updated weighted average and
the Lee-Yang’s prediction (1960) on the basis of no second-class currents~\cite{theory_LY}. 
    }
\label{fig_comparison}
\end{figure}

In summary, using $(10087\pm44) \times 10^6$ $J/\psi$ decay events
collected with the BESIII detector at $\sqrt{s}=3.097$ GeV, the
semileptonic hyperon decay $\Sigma^+ \to \Lambda e^+ \nu_{e}$ is
studied at a collider experiment for the first time.  This is also the
first experimental study after a more than 50-year hiatus.  Based on
the DT method, we perform the first direct measurement of the absolute
branching fraction of $\Sigma^+ \to \Lambda e^+ \nu_{e}$ to be
${\mathcal B}(\Sigma^+ \to \Lambda e^+ \nu_{e}) = [2.93\pm0.74(\rm
  stat) \pm 0.13(\rm syst)]\times 10^{-5}$, which is consistent with
all the previous indirect measurements within uncertainty and is the
most precise result (about 25\% improvement in precision) obtained in
a single experiment to date~\cite{pdg2022}.  Combining with the
well-measured branching fraction of $\Sigma^- \to \Lambda e^-
\bar{\nu}_{e}$ and the lifetimes of $\Sigma^{\pm}$, we determine the
$R$ value defined in Eq.~\ref{eq_r} to be $R = 1.06 \pm 0.28$.
Fig.~\ref{fig_comparison} shows the comparison between our $R$ value and the SM prediction of Ref.~\cite{theory_LY}, together with values from previous fixed-target experiments~\cite{old_BF_result_1,old_BF_result_2,old_BF_result_3}
recalculated with the latest PDG values of ${\mathcal B}(\Sigma^- \to \Lambda e^- \bar{\nu}_{e})$ and $\tau_{\Sigma^{\pm}}$;
we present the updated weighted average $R$ value, including our measurement,  as $1.37 \pm 0.25$.
This Letter contributes to the most precise measurement of $R$ in a single experiment;
this result is within 1.8 standard deviations of the theoretical
 calculation~\cite{theory_LY} in the absence of second-class
    currents.

\acknowledgments
The BESIII collaboration thanks the staff of BEPCII and the IHEP computing center for their strong support. This work is supported in part by National Key R\&D Program of China under Contracts Nos. 2020YFA0406300, 2020YFA0406400; National Natural Science Foundation of China (NSFC) under Contracts Nos. 11805037, 11635010, 11735014, 11835012, 11935015, 11935016, 11935018, 11961141012, 12022510, 12025502, 12035009, 12035013, 12061131003, 12192260, 12192261, 12192262, 12192263, 12192264, 12192265; the Chinese Academy of Sciences (CAS) Large-Scale Scientific Facility Program; the CAS Center for Excellence in Particle Physics (CCEPP); Joint Large-Scale Scientific Facility Funds of the NSFC and CAS under Contract Nos. U1832121, U1832207; CAS Key Research Program of Frontier Sciences under Contracts Nos. QYZDJ-SSW-SLH003, QYZDJ-SSW-SLH040; 100 Talents Program of CAS; The Institute of Nuclear and Particle Physics (INPAC) and Shanghai Key Laboratory for Particle Physics and Cosmology; ERC under Contract No. 758462; European Union's Horizon 2020 research and innovation programme under Marie Sklodowska-Curie grant agreement under Contract No. 894790; German Research Foundation DFG under Contracts Nos. 443159800, 455635585, Collaborative Research Center CRC 1044, FOR5327, GRK 2149; Istituto Nazionale di Fisica Nucleare, Italy; Ministry of Development of Turkey under Contract No. DPT2006K-120470; National Science and Technology fund; National Science Research and Innovation Fund (NSRF) via the Program Management Unit for Human Resources \& Institutional Development, Research and Innovation under Contract No. B16F640076; Olle Engkvist Foundation under Contract No. 200-0605; Polish National Science Centre under Contract No. 2019/35/O/ST2/02907; STFC (United Kingdom); Suranaree University of Technology (SUT), Thailand Science Research and Innovation (TSRI), and National Science Research and Innovation Fund (NSRF) under Contract No. 160355; The Royal Society, UK under Contracts Nos. DH140054, DH160214; The Swedish Research Council; U. S. Department of Energy under Contract No. DE-FG02-05ER41374.



\begin{thebibliography}{**}
\bibitem{indirect_m}
In the previous measurements of ${\mathcal B}(\Sigma^+ \to \Lambda e^+ \nu_{e})$~\cite{old_BF_result_1, old_BF_result_2, old_BF_result_3}, 
in order to get the branching fraction, they had to determine the total number of produced $\Sigma^+$ at first.
In Refs.~\cite{old_BF_result_1, old_BF_result_2}, they deduced the number of produced $\Sigma^+$ using
the production rate of $\frac{\Sigma^+}{\Sigma^-}$.
In Ref.~\cite{old_BF_result_3}, they determined the number of produced $\Sigma^+$ by counting the number of $\Sigma^+$'s 
decaying into protons.
\end{thebibliography}
\end{document}